\documentclass{moriond}

\def\Journal#1#2#3#4{{#1} {\bf #2}, #3 (#4)}

\def\PRD{{\em Phys. Rev.} D}

\def\be{\begin{equation}}
\def\ee{\end{equation}}
\def\bea{\begin{eqnarray}}
\def\eea{\end{eqnarray}}

\newcommand{\Photo}{\includegraphics[height=35mm]{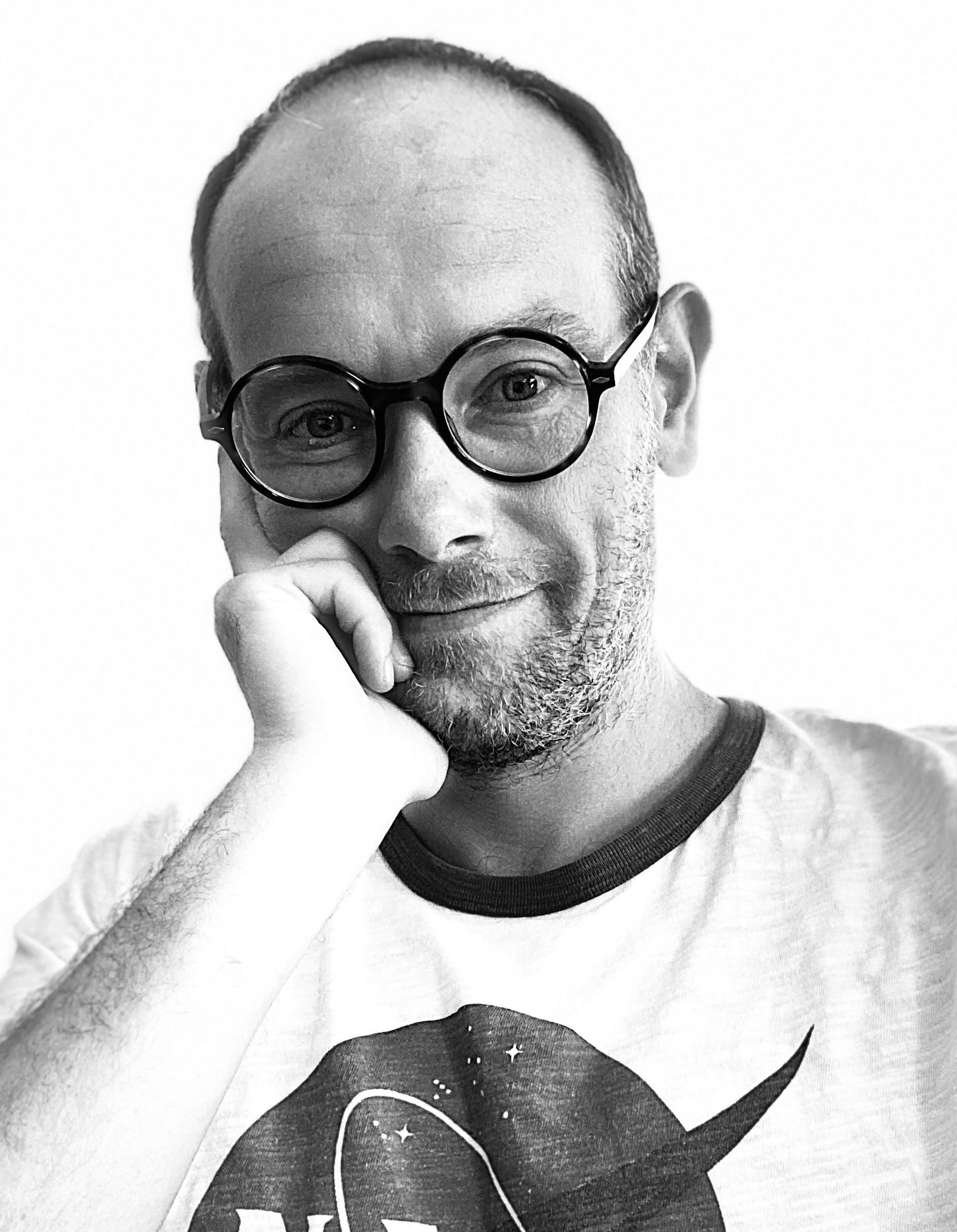}}

\begin{document}
\vspace*{4cm}
\title{Black-holes and neutron stars in entangled relativity}

\author{Olivier Minazzoli}

\address{Artemis, Universit\'e C\^ote d'Azur, CNRS, Observatoire C\^ote d'Azur\\ BP4229, 06304, Nice Cedex 4, France}

\maketitle\abstracts{
I present two results that show that, despite its unusual non-linear form, the phenomenology of entangled relativity remains close to the one of general relativity---without having any free parameter that can be fine tuned in order to facilitate this. In particular, I present the analytical solutions for spherically charged black-holes, and both the analytical and numerical solutions for neutron stars. 
}

\section{Introduction}

Entangled relativity is a general theory of relativity that has a (very) specific non-linear coupling between matter and curvature at the level of the action, which prevents the definition of the whole theory without defining the matter content of the theory in the first place---hence, matter and curvature are \textit{entangled}, in the etymological sense, as one cannot be considered without the other. Indeed, the whole Lagrangian density of the theory reads $\mathcal{L}=-\xi/2 ~\mathcal{L}_m^2/R$ \cite{minazzoli:2018pr}, where $\xi$ is a dimensionfull constant, $\mathcal{L}_m$ is the Lagrangian density of matter (e.g. the standard model of particle physics) and $R$ the usual Ricci scalar. This has the nice outcome that the theory does not obviously violate the \textit{principle of relativity of inertia} that Einstein named \textit{Mach's principle}. Indeed, although it has been somewhat forgotten, many people over the last century (including the young Einstein) \cite{mach} had the firm belief that a consistent theory of relativity had to satisfy the \textit{principle of relativity of inertia} that, roughly speaking, states that \textit{inertia} must be defined with respect to surrounding matter, and not with respect to some space (or spacetime) given beforehand---\textit{\`a la Newton}.
In other words, a consistent general theory of relativity should not rely on any absolute structure.
As a consequence, Einstein (and others) believed that the metric---from which inertia is defined in relativistic theories---had to be entirely determined by matter,\footnote{If spacetime would not \textit{entirely} be determined by matter, then it would only be \textit{partially} determined by matter---which is less satisfying from a philosophical perspective.} such that a satisfying theory of relativity would not allow for the existence of vacuum solutions. Unfortunately, general relativity possesses vacuum solutions---as do most (if not all) alternatives to general relativity---such that it blatantly violates the \textit{principle of relativity of inertia}. Incidentally, another thing that seems to have been somewhat forgotten is that this fact precisely is what led Einstein to add the cosmological constant to the equation of general relativity.\footnote{See \href{https://einsteinpapers.press.princeton.edu/vol7-trans/52}{https://einsteinpapers.press.princeton.edu/vol7-trans/52} and \href{https://einsteinpapers.press.princeton.edu/vol6-trans/433}{https://einsteinpapers.press.princeton.edu/vol6-trans/433}.}

But entangled relativity would not be interesting if it only satisfied \textit{Mach's principle}. What is compelling about this theory is that---despite its weird looking non-linear Lagrangian density---it leads to a phenomenology in various situations that is close (and even sometimes equivalent) to the one of general relativity---even though it has no free parameter to adjust at the classical level.\footnote{This is actually quite astonishing if one considers the inflation of free parameters in the literature---not even talking about the string landscape.} Indeed, the only parameter of the theory ($\xi$) has no impact on the classical phenomenology of the theory. It is, therefore, a purely quantum parameter, which is related to the quantum of action of the theory \cite{minazzoli:2022}.

It turns out that the nice features of entangled relativity follow from the fact that its Lagrangian density respects the following equality
\begin{equation}
-\frac{\xi}{2} \frac{\mathcal{L}_m^2}{R} \equiv \frac{\xi}{\kappa} \left(\frac{R}{2\kappa}+\mathcal{L}_m \right),\label{eq:equality}
\end{equation}
where $\kappa:=-R/\mathcal{L}_m$ is promoted to a scalar-field with respect to general relativity (for which $\kappa = -R/T$ instead). This means that in entangled relativity, the amplitude with which spacetime is curved by matter depends on the localisation. This improbable equality (\ref{eq:equality}) can actually be shown at the simple algebraic level already (noting that $-\xi/2~\mathcal{L}_m^2/R =-\xi/2( a~\mathcal{L}_m^2/R^2~R+b~\mathcal{L}_m/R ~\mathcal{L}_m)$ with $a+b=1$ and $b/2=1$). Otherwise, one can also check that all the field equations that derive from the two Lagrangian densities are the same.

In this what follows, I will present two aspects of its phenomenology: sphericaly charged black-holes and neutron stars.

\section{Spherically charged black-holes}

The specific non-linear nature of entangled relativity forbids one to consider spacetime without defining matter in the first place. Hence, even black-holes have to be derived after assuming some sort of material field. Naturally, the first non-vacuum black-hole that we considered were spherically symetric charged black-holes \cite{minazzoli:2021ej}. The metric for such balck-holes reads

\begin{equation}
\mathrm{d} s^{2}=-\lambda_{0}^{2} \mathrm{~d} t^{2}+\lambda_{r}^{-2} \mathrm{~d} r^{2}+\rho^{2}\left(\mathrm{~d} \theta^{2}+\sin ^{2} \theta \mathrm{d} \varphi^{2}\right),
\end{equation}
with
\begin{equation}
\begin{array}{l}
\lambda_{0}^{2}=\left(1-\frac{r_{+}}{r}\right)\left(1-\frac{r_{-}}{r}\right)^{15 / 13}, \\
\lambda_{r}^{2}=\left(1-\frac{r_{+}}{r}\right)\left(1-\frac{r_{-}}{r}\right)^{7 / 13}, \\
\rho^{2}=r^{2}\left(1-\frac{r_{-}}{r}\right)^{6 / 13},
\end{array}
\end{equation}
where the mass $M$ and the charge $Q$ are related to the $r_+$ and $r_-$ parameters as follows
\begin{equation}
2 M=r_{+}+\left(\frac{1-a^{2}}{1+a^{2}}\right) r_{-},
\end{equation}
and
\begin{equation}
Q^{2}=\frac{r_{-} r_{+}}{1+a^{2}}.
\end{equation}
The limit $r_- \rightarrow 0$ corresponds to the Schwarzschild black-holes, whereas the limit $r_+ \rightarrow 0$ corresponds to black-holes with scalar hair---where $a$ parametrizes the amount of scalar hair. However, the latter cannot be formed by gravitational collapse, as scalar hair are known to be generically radiated away during a collapse into a black-hole. Hence, it seems that for astrophysical conditions---that is, $Q\sim 0$---Schwarzschild black-holes are good enough approximations of spherical black-holes in entangled relativity. We conjectured that it is a general feature of the black-hole solutions of general relativity \cite{minazzoli:2021ej}.
\section{Neutron stars}

\subsection{Analytical solutions}

In the Einstein frame,\footnote{That is, after a conformal transformation of the metric that is such that the Ricci scalar based upon the new metric seems to be minimally coupled in the Lagragian density.} a general solution that is valid outside any spherical compact object is \cite{arruga:2021ep}
\begin{eqnarray}
d s^{2}=&-\left(1-\frac{2 m}{\beta r}\right)^{\beta} \mathrm{d} t^{2}+\left(1-\frac{2 m}{\beta r}\right)^{-\beta} \mathrm{d} r^{2} +r^{2}\left(1-\frac{2 m}{\beta r}\right)^{1-\beta}\left[\mathrm{d} \theta^{2}+\sin ^{2} \theta \mathrm{d} \psi^{2}\right],
\end{eqnarray}
where $\beta \in ]0;1]$ parametrizes the amount of scalar charge of the object. This solution is obviously somewhat related to the limit $r_+ \rightarrow 0$ of the metric of the last subsection. As we will see in the next sub-section, $\beta$ is surprisingly close to unity (even) for neutron stars in entangled relativity.
\subsection{Numerical solutions}

\begin{figure} 
\centerline{\includegraphics[width=0.55\linewidth]{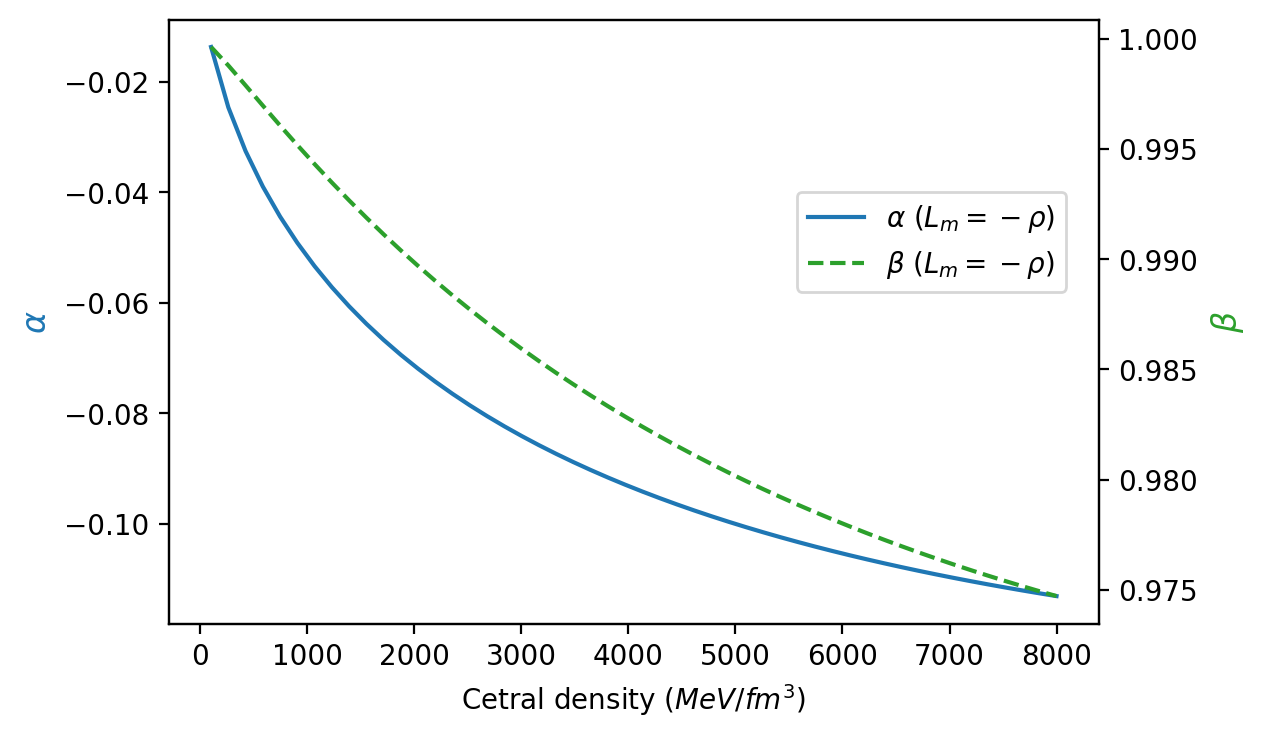}}
\caption[]{Values of the parameters $\alpha$ and $\beta$ with respect to the central density of the compact object when assuming $\mathcal{L}_m=-\rho$, with $\alpha^2 = (1-\beta)/(1+\beta)$.}
\label{fig:beta}
\end{figure}

\begin{figure} 
\centerline{\includegraphics[width=0.55\linewidth]{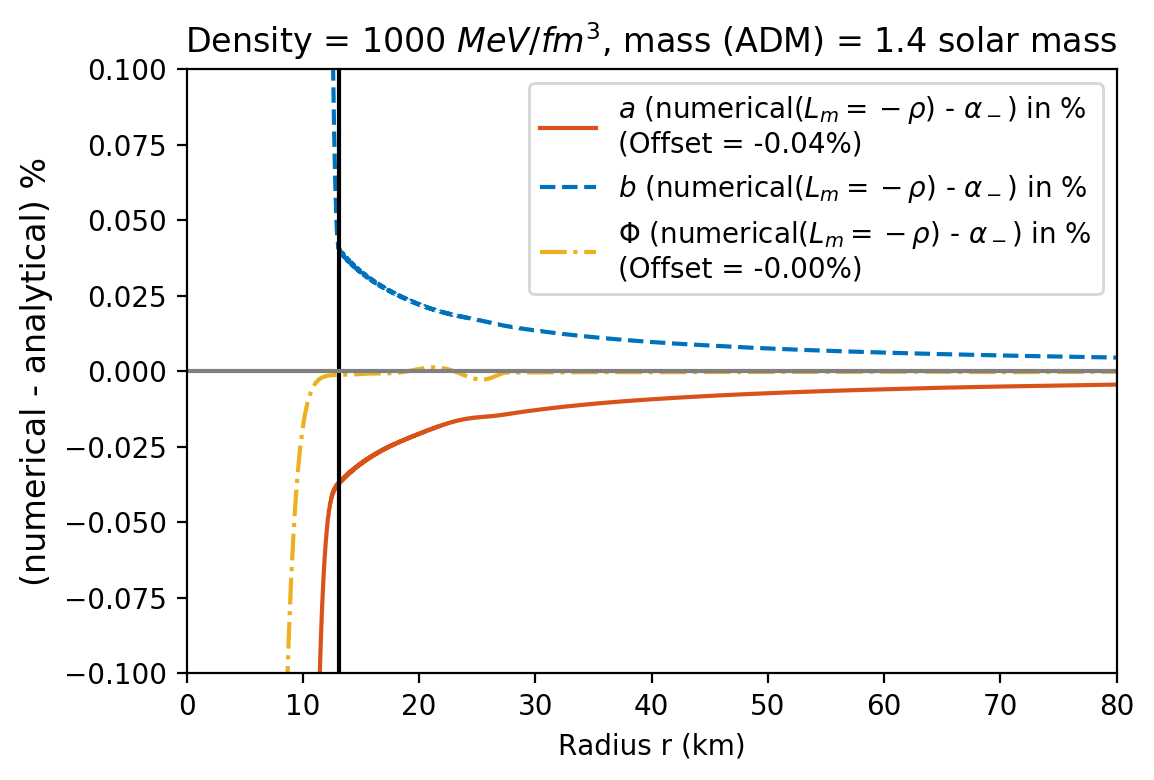}}
\caption[]{Relative differences between the analitical and numerical solutions for a given simulation. $a$ is the time-time component of the metric, $b$ the space-space component, and $\Phi$ the scalar degree of freedom that comes from the non-linear nature of the Lagrangian density of the theory. The vertical line represents the radius of the compact object in the simulation. A sub-permille offset appears due to the fact that a perfect matching of the mass in each solution (i.e. numerical versus analytical) should be done at infinity, while we have done it at the maximal radius of our simulation $R_l=10000$km for obvious practical reasons.}
\label{fig:diff}
\end{figure}

Alternatively, we numerically solved the Tolman-Oppenheimer–Volkoff equations \cite{arruga:2021pr}. An ambiguity however arises at the level of the field equations. Indeed, different possible values of the on-shell matter Lagrangian density give different results. Either $\mathcal{L}_m = T$ on-shell, and one recovers exactly the results of general relativity, or $\mathcal{L}_m  = -\rho$ on-shell---where $\rho$ is the total energy density---and one gets the values of $\beta$ given in Figure \ref{fig:beta}. What is striking is that even in the worst case scenario, $\beta$ remains only a few percent away from unity. Of course, for the messy fluids that make up a neutron star---which involve many different physical phenomena at various scales---one might very well have a mixture between the two possibilities \cite{arruga:2021pr}. Determining the on-shell matter Lagrangian density from first principles remains a challenge in entangled relativity at the time of this writing.
The numerical and the analytical solutions match perfectly well outside the compact object, as one can see in Figure \ref{fig:diff}.

\section{Conclusion}

Despite its very unusual form, entangled relativity might be consistent with the laws of physics, given that it possesses general relativity as a limit in various (classical) situations. What makes the theory remarkable with respect to other alternative theories to general relativity is what follows
\begin{itemize}
\item It does not have any free parameter at the classical level.
\item It seems to satisfy the three principles that Einstein posited for a satisfying general theory of relativity---that is, the theory is covariant (\textit{the principle of relativity}), the metric tensor encodes the mechanical properties of space as well as the inertia of bodies and gravitation (\textit{the equivalence principle}), and the metric is entirely determined by matter (\textit{the principle of relativity of inertia}, or \textit{Mach's principle}).
\footnote{See \href{https://einsteinpapers.press.princeton.edu/vol7-trans/49}{https://einsteinpapers.press.princeton.edu/vol7-trans/49}.} 
\end{itemize}
Although note that, the \textit{equivalence principle} in the sense of the (known as) \textit{Einstein equivalence principle} \cite{will:2014lr} is only realized approximatively in entangled relativity. The amplitude of the violation of the \textit{Einstein equivalence principle} is nonetheless very weak, but varies depending on the on-shell values of the Lagrangian density for realistic gravitating bodies. As long as the on-shell value of the matter Lagrangian density for cellestial bodies cannot be determined from first principles, one cannot rigorously estimate the amplitude of the violation of the \textit{Einstein equivalence principle} expected from this theory, nor the general deviations from general relativity, for that matter. Fortunately, the on-shell matter Lagrangian for electromagnetic fields (in near vacuum situations otherwise) can easily be computed, such that they might be the best means to probe the theory with experiments and observations.


\section*{References}

\begin{thebibliography}{99}
\bibitem{minazzoli:2018pr} O. Minazzoli, \Journal{\PRD}{98}{124020}{2018}
\bibitem{mach} Mach's Principle: From Newton's Bucket to Quantum Gravity, ed. J. Barbour and H. Pfister, {\em Einstein studies Vol.6}
(Birkh\"auser, Boston, 1995).
\bibitem{minazzoli:2022} O. Minazzoli, {\em to appear soon}
\bibitem{minazzoli:2021ej} O. Minazzoli and E. Santos, \Journal{European Physical Journal C}{81}{640}{2021}
\bibitem{arruga:2021ep} D. Arruga and O. Minazzoli, \Journal{European Physical Journal C}{81}{1007}{2021}
\bibitem{arruga:2021pr} D. Arruga, O. Rousselle and O. Minazzoli, \Journal{\PRD}{103}{024034}{2021}
\bibitem{will:2014lr} C. Will, \Journal{Living Reviews in Relativity}{17}{4}{2014}
%
%
%

\end{thebibliography}
\end{document}